\documentclass[12pt]{article}
\usepackage{amsmath}
\usepackage{graphicx}
\usepackage{natbib}
\usepackage{url} 

\usepackage{tikz}
\usetikzlibrary{shapes}
\usepackage{xcolor}

\usepackage{textcomp}
\newcommand{\textapprox}{\raisebox{0.5ex}{\texttildelow}}

\makeatletter
\newcommand{\tpmod}[1]{{\@displayfalse\pmod{#1}}}
\makeatother

\DeclareMathSizes{12}{12}{10}{10}

\newcommand{\blind}{0}

\begin{document}

\def\spacingset#1{\renewcommand{\baselinestretch}%
{#1}\small\normalsize} \spacingset{1}


\if0\blind
{
  \title{\bf Non-Uniform Gaussian Blur of Hexagonal Bins in Cartesian Coordinates}
  \author{Reinier Vleugels\thanks{
    The first author is an MSc student at the Department of Computer Science, IBIVU Centre for Integrative Bioinformatics, Vrije Universiteit Amsterdam. This work was conducted as part of his thesis.}\hspace{.2cm}\\
    Department of Computer Science, \\
    IBIVU Centre for Integrative Bioinformatics,\\
Vrije Universiteit Amsterdam\\
    and \\
    Magnus Palmblad \\
    Center for Proteomics and Metabolomics, \\
    Leiden University Medical Center}
  \maketitle
} \fi

\if1\blind
{
  \bigskip
  \bigskip
  \bigskip
  \begin{center}
    {\LARGE\bf Title}
\end{center}
  \medskip
} \fi

\bigskip
\begin{abstract}
In a recent application of the Bokeh Python library for visualizing physico-chemical properties of chemical entities text-mined from the scientific literature, we found ourselves facing the task of smoothing hexagonally binned data in Cartesian coordinates. To the best of our knowledge, no documentation for how to do this exist in the public domain. This short paper  shows how to accomplish this in general and for Bokeh in particular. We illustrate the method with a real-world example and discuss some potential advantages of using hexagonal bins in these and similar applications.
\end{abstract}

\noindent%
\vfill

\newpage
\spacingset{1.5} 
\section{Introduction}
\label{sec:intro}

Hexagonal binning is a popular alternative for creating two-dimensional histograms that captures most shapes better than the more common rectilinear binning methods \citep{carr87}. Hexagons are more similar to circles than squares which means data is aggregated more tightly around the bin center. The minimum distance from a bin center to the nearest bin border is \textapprox7.5\% longer in hexagonal bins than in equiareal square bins, which likely improves accuracy of bin selection when interacting with histograms using a computer mouse or touchscreen. The major disadvantage is the added complexity of computation and visualization. It is also impossible to exactly subdivide or aggregate hexagonal bins into smaller or larger hexagons, which is easy to do with rectangular bins. However, in some applications, these drawbacks are not critical.

One of the added complexities when working with hexagonal bins is the coordinate system and transformations between the hexagonal and Cartesian coordinate systems. There are several hexagonal grid systems, such as the \textit{axial} or \textit{trapezoidal coordinates} used by the \textit{Bokeh} Python library \citep{bokeh}, the \textit{offset coordinates} used by the \textit{TikZ Shapes} \LaTeX\ library (also to generate Figures 1-5 in this paper) and \textit{cube coordinates} with three axes. Other software, including the R \textit{tess} package \citep{hohna13}, use a one-dimensional offset enumeration to address individual hexagons. In this paper we will focus on offset and axial coordinates, though any hexagonal coordinates or numbering system can be transformed into Cartesian coordinates.

It is sometimes advantageous to smooth images or histograms to reduce noise and emphasize features of interest. Blurring allows data visualization at a higher resolution without requiring a large numbers of counts in each bin. Blurring can also be used to represent uncertainly in measurements or predictions. When the hexagonal tiles represent spatial data such as a map or image, blurring can be done in the hexagonal coordinate system, with the distance between two tiles being the minimum number of steps between them. Each hexagon thus have 6 neighboring tiles with distance 1, 12 tiles of distance 2, 18 of distance 3, etc - in general $6n$ tiles in the $n$-th shell around a given tile (Figure 1). Blurs based on this distance metric are approximately radial.

\begin{figure}[ht]
\begin{center}
\begin{tikzpicture} [hexa/.style= {shape=regular polygon,
                                   regular polygon sides=6,
                                   minimum size=2cm, draw,
                                   inner sep=0,anchor=south,
                                   fill=none}]

\node[hexa] (h1;1) at ({0},{0}) {(0,0)};

\node[hexa] (h1;2) at ({0},{2*sin(60)}) {(0,1)};
\node[hexa] (h1;3) at ({0},{4*sin(60)}) {(0,2)};

\node[hexa] (h2;1) at ({1.5},{1*sin(60)}) {(1,0)};
\node[hexa] (h2;2) at ({1.5},{3*sin(60)}) {(1,1)};

\node[hexa] (h3;2) at ({3},{2*sin(60)}) {(2,1)};
\node[hexa] (h3;1) at ({3},{0}) {(2,0)};

\node[hexa] (h1;2) at ({0},{-2*sin(60)}) {(0,-1)};
\node[hexa] (h1;3) at ({0},{-4*sin(60)}) {(0,-2)};

\node[hexa] (h2;1) at ({1.5},{-1*sin(60)}) {(1,-1)};
\node[hexa] (h2;2) at ({1.5},{-3*sin(60)}) {(1,-2)};

\node[hexa] (h3;2) at ({3},{-2*sin(60)}) {(2,-1)};

\node[hexa] (h2;1) at ({-1.5},{-1*sin(60)}) {(-1,-1)};
\node[hexa] (h2;2) at ({-1.5},{-3*sin(60)}) {(-1,-2)};

\node[hexa] (h3;2) at ({-3},{-2*sin(60)}) {(-2,-1)};

\node[hexa] (h2;1) at ({-1.5},{1*sin(60)}) {(-1,0)};
\node[hexa] (h2;2) at ({-1.5},{3*sin(60)}) {(-1,1)};

\node[hexa] (h3;2) at ({-3},{2*sin(60)}) {(-2,1)};
\node[hexa] (h3;1) at ({-3},{0}) {(-2,0)};

\draw[->,dotted] (0,{-5*sin(60)}) -- (0,{7*sin(60)});
\draw[->,dotted] (-4.5-0.25, {sin(60)*1.5}) -- (4.5+0.25, {sin(60)*1.5});

\node[] at (4.5+0.25,{sin(60)*1.5-0.3}) {$x$};
\node[] at (0.3,3.5*1.732) {$y$};

\end{tikzpicture}
\caption{Hexagonal tiles with offset hexagonal coordinates $(x,y)$ in two shells around (0,0). The $x$ axis is here drawn in the middle of the tiles with $y=0$.}
\label{fig1}
\end{center}
\end{figure}
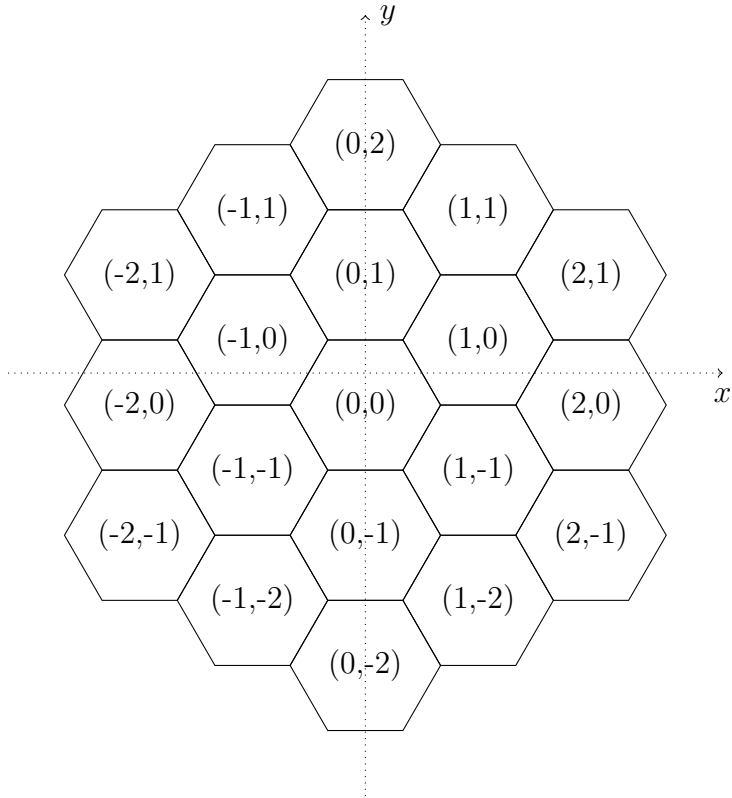

\begin{figure}[ht]
\begin{center}
\begin{tikzpicture} [hexa/.style= {shape=regular polygon,
                                   regular polygon sides=6,
                                   minimum size=2cm, draw,
                                   inner sep=0,anchor=south,
                                   fill=none}]

\node[hexa] (h1;1) at ({0},{0}) {(0,0)};

\node[hexa] (h1;2) at ({0},{2*sin(60)}) {(0,-1)};
\node[hexa] (h1;3) at ({0},{4*sin(60)}) {(0,-2)};

\node[hexa] (h2;1) at ({1.5},{1*sin(60)}) {(1,-1)};
\node[hexa] (h2;2) at ({1.5},{3*sin(60)}) {(1,-2)};

\node[hexa] (h3;2) at ({3},{2*sin(60)}) {(2,-2)};
\node[hexa] (h3;1) at ({3},{0}) {(2,-1)};

\node[hexa] (h1;2) at ({0},{-2*sin(60)}) {(0,1)};
\node[hexa] (h1;3) at ({0},{-4*sin(60)}) {(0,2)};

\node[hexa] (h2;1) at ({1.5},{-1*sin(60)}) {(1,0)}; 
\node[hexa] (h2;2) at ({1.5},{-3*sin(60)}) {(1,1)};

\node[hexa] (h3;2) at ({3},{-2*sin(60)}) {(2,0)};

\node[hexa] (h2;1) at ({-1.5},{-1*sin(60)}) {(-1,1)};
\node[hexa] (h2;2) at ({-1.5},{-3*sin(60)}) {(-1,2)};

\node[hexa] (h3;2) at ({-3},{-2*sin(60)}) {(-2,2)};

\node[hexa] (h2;1) at ({-1.5},{1*sin(60)}) {(-1,0)};
\node[hexa] (h2;2) at ({-1.5},{3*sin(60)}) {(-1,-1)};

\node[hexa] (h3;2) at ({-3},{2*sin(60)}) {(-2,0)};
\node[hexa] (h3;1) at ({-3},{0}) {(-2,1)};

\draw[->,dotted] (0,3.5*1.732) -- (0,-2.5*1.732);
\draw[->,dotted] (-4.5,2*1.732) -- (4.5,-1.732);

\node[] at (4.5-0.15,-1.732-0.3) {$q$};
\node[] at (0.3,-2.5*1.732) {$r$};

\end{tikzpicture}
\caption{Hexagonal tiles with axial (trapezoidal) coordinates $(q,r)$ in the same two shells around (0,0). This is the hexagonal coordinate system used by Bokeh.}
\label{fig2}
\end{center}
\end{figure}
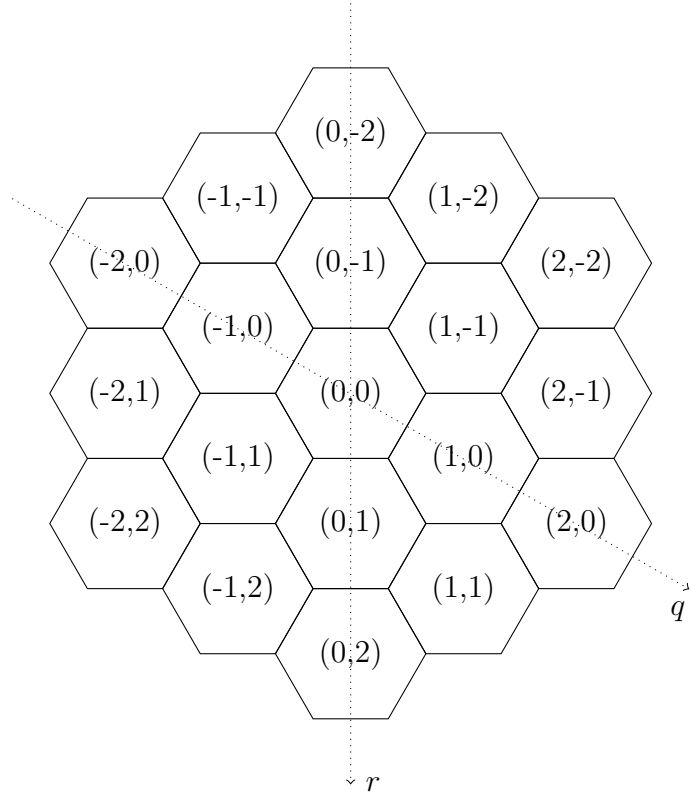

However, in one of our use cases, based on previous work \citep{palmblad19}, the two dimensions do not represent spatial locations as in an map or image, but two different physical variables, with different underlying uncertainties. To smooth the histogram and visualize these uncertainties, we have to use the original physical dimensions or coordinates. As we still wish to use hexagonal bins, for the reasons mentioned above, we have to evaluate the smoothing function in the Cartesian coordinates of the bins, e.g. the bin centers. Note that we do not wish to smooth the \textit{image}. The hexagonal bins should remain distinct to allow the user to interact with the binned data. Here we show how this can be done, and demonstrate the result using Bokeh. For a more comprehensive and interactive overview of the many different hexagonal grid systems, see Amit Patel's excellent blog post on the subject \citep{patel13}.

\pagebreak
\section{Methods}
\label{sec:meth}

A hexagon can be seen as made up by six equilateral triangles. From elementary trigonometry, we know the ratio of the height to the side in an equilateral triangle is $\sin{60^{\circ}}$ or $\sqrt{3}/2$. If we define the length of the sides of the regular hexagons as 1, the center-to-center distance between two adjacent hexagons, that is two hexagons sharing one edge, is therefore $\sqrt{3}$. Between hexagonal bins in offset rows or columns, the distance is 3/2 in one dimension and $\sqrt{3}/2$ in the other (Figure 2). The transformations from offset hexagonal ($H$) coordinates (Figure 1) to Cartesian ($C$) ones are thus:

\begin{equation}
\quad
\begin{bmatrix}
x \\
y
\end{bmatrix}_{C}
=
\begin{cases}    
    \begin{bmatrix}
3/2 & 0 \\
0 & \sqrt{3}
\end{bmatrix}
\begin{bmatrix}
x \\
y
\end{bmatrix}_{H} & \text{if } x_{H} \equiv 0\tpmod{2}\\
\begin{bmatrix}
3/2 & 0 \\
0 & \sqrt{3}
\end{bmatrix}
\begin{bmatrix}
x \\
y
\end{bmatrix}_{H}
+
\begin{bmatrix}
0 \\
1/2
\end{bmatrix}_{C}
& \text{if } x_{H} \equiv 1 \tpmod{2}
  \end{cases}
\end{equation}

\vspace{12pt}
With an angle $60^{\circ}$ between the axes, the transformation from the hexagonal coordinates $q$ and $r$ in Figure 2 to Cartesian coordinates is independent of coordinate parity:

\begin{equation}
\quad
\begin{bmatrix}
x \\
y
\end{bmatrix}_{C}
=
\begin{bmatrix}
3/2 & 0 \\
-\sqrt{3}/2 & -\sqrt{3}
\end{bmatrix}
\begin{bmatrix}
q \\
r
\end{bmatrix}_{H}
\quad
\end{equation}

\vspace{12pt}
This transformation is illustrated in Figure 3 below. The Cartesian coordinates here refer to the centers of the hexagonal bins:

\vspace{12pt}
\begin{figure}[ht]
\begin{center}
\begin{tikzpicture} [hexa/.style= {shape=regular polygon,
                                   regular polygon sides=6,
                                   minimum size=2cm, draw,
                                   inner sep=0,anchor=south,
                                   fill=none}]

\node[hexa] (h1;1) at ({0},{0}) {(0,0)};
\node[hexa] (h1;2) at ({0},{2*sin(60)}) {(0,-1)};
\node[hexa] (h2;1) at ({1.5},{1*sin(60)}) {(1,-1)};
\node[hexa] (h3;1) at ({3},{0}) {(2,-1)};

\node[hexa] (h1;1) at ({7+2/3},{0}) {(0,0)};
\node[hexa] (h1;2) at ({7+2/3},{2*sin(60)}) {(0,$\sqrt{3}$)};
\node[hexa] (h2;1) at ({9+1/6},{1*sin(60)}) {($\frac{3}{2}$,$\frac{\sqrt{3}}{2}$)};
\node[hexa] (h3;1) at ({10+2/3},{0}) {(3,0)};

\draw[<->] (6,0.866025) -- (6,2.598076);
\draw[dotted] (6,0.866025) -- (6.6,0.866025);
\draw[dotted] (6,2.598076) -- (6.6,2.598076);
\node[] at (5.5,1.732050) {$\sqrt{3}$};
\node[] at (7.67,-0.3) {1};

\end{tikzpicture}
\caption{Four hexagonal tiles in axial hexagonal coordinates as in Figure 2 (left) and in corresponding Cartesian coordinates in units of hexagon side length (right). The Cartesian coordinates corresponding to $(q,r)$ in axial coordinates are $(3q/2,\sqrt{3}(q/2-r))$. The Cartesian coordinates corresponding to $(x,y)$ in offset hexagonal coordinates (Figure 1) are $(3x/2,\sqrt{3}y)$ for even columns ($x\mod{}2=0$) and $(3x/2,\sqrt{3}y+\sqrt{3}/2)$ for odd ($x\mod{}2=1$).}
\label{fig3}
\end{center}
\end{figure}
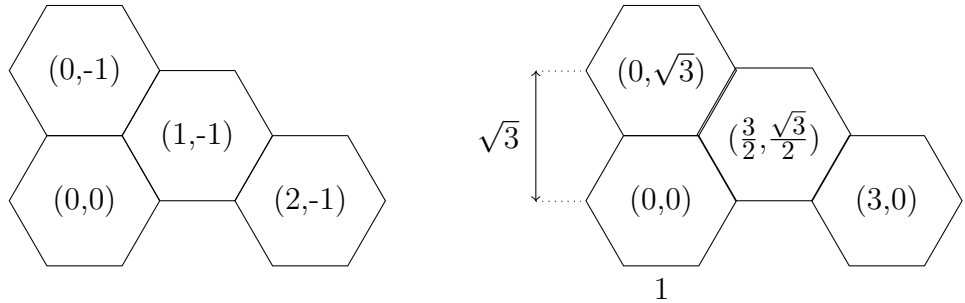

\pagebreak

As the two Cartesian dimensions in our use case represent different physical dimensions with different uncertainties of measurement or estimation, the smoothing should not have rotational symmetry like a radial blur. If the errors are independent, the Gaussian kernel or signal $S$ that should be transferred from $(0,0)$ to $(x,y)$ in Cartesian coordinates is:

\begin{equation}
\vspace{12pt}
S(x,y) = \frac{1}{{\sigma_x \sigma_y \sqrt {2\pi } }} e^{{{ -  {x} ^2 {y} ^2 } \mathord{\left/ {\vphantom {{ - \left( {x} \right)^2 }  {2\sigma_x ^2 }{2\sigma_y ^2 }}} \right. \kern-\nulldelimiterspace} {2\sigma_x ^2 } {\sigma_y ^2 }}    }
\vspace{12pt}
\end{equation}

where $\sigma_x$ and $\sigma_y$ are the standard deviations in the Cartesian coordinates corresponding to the two physical dimensions. We only need to evaluate $S(x,y)$ once for any pair of $\sigma_x$ and $\sigma_y$ and in practise only for a limited number hexagonal bins (Figure 4):

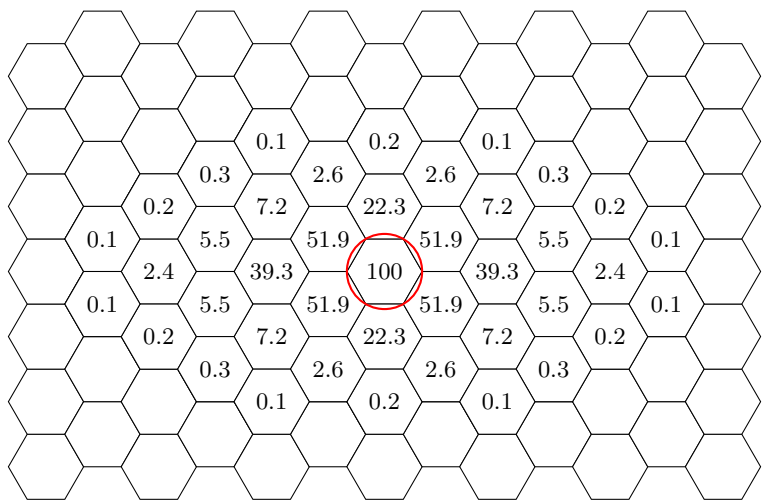
\begin{figure}[ht]
\begin{center}
\begin{tikzpicture} [hexa/.style= {shape=regular polygon,
                                   regular polygon sides=6,
                                   minimum size=1cm, draw,
                                   inner sep=0,anchor=south,
                                   fill=white}]
\fontsize{9}{11} 
\foreach \j in {0,...,12}{ 
    \ifodd\j 
       \foreach \i in {1,...,7}{\node[hexa] (h\j;\i) at ({\j/2+\j/4},{(\i+1/2)*sin(60)}){};}        
    \else
        \foreach \i in {1,...,7}{\node[hexa] (h\j;\i) at ({\j/2+\j/4},{\i*sin(60)}){};}
    \fi}


\node[hexa] (h6;4) at ({6/2+6/4},{4*sin(60)}) {100};

\node[hexa] (h4;4) at ({4/2+4/4},{4*sin(60)}) {39.3};
\node[hexa] (h8;4) at ({8/2+8/4},{4*sin(60)}) {39.3};

\node[hexa] (h2;4) at ({2/2+2/4},{4*sin(60)}) {2.4};
\node[hexa] (h10;4) at ({10/2+10/4},{4*sin(60)}) {2.4};

\node[hexa] (h6;3) at ({6/2+6/4},{3*sin(60)}) {22.3};
\node[hexa] (h6;5) at ({6/2+6/4},{5*sin(60)}) {22.3};

\node[hexa] (h6;2) at ({6/2+6/4},{2*sin(60)}) {0.2};
\node[hexa] (h6;6) at ({6/2+6/4},{6*sin(60)}) {0.2};

\node[hexa] (h4;3) at ({4/2+4/4},{3*sin(60)}) {7.2};
\node[hexa] (h4;5) at ({4/2+4/4},{5*sin(60)}) {7.2};
\node[hexa] (h8;3) at ({8/2+8/4},{3*sin(60)}) {7.2};
\node[hexa] (h8;5) at ({8/2+8/4},{5*sin(60)}) {7.2};

\node[hexa] (h2;3) at ({2/2+2/4},{3*sin(60)}) {0.2};
\node[hexa] (h10;3) at ({10/2+10/4},{3*sin(60)}) {0.2};
\node[hexa] (h2;5) at ({2/2+2/4},{5*sin(60)}) {0.2};
\node[hexa] (h10;5) at ({10/2+10/4},{5*sin(60)}) {0.2};

\node[hexa] (h5;3) at ({5/2+5/4},{(3+1/2)*sin(60)}) {51.9};
\node[hexa] (h5;4) at ({5/2+5/4},{(4+1/2)*sin(60)}) {51.9};
\node[hexa] (h7;3) at ({7/2+7/4},{(3+1/2)*sin(60)}) {51.9};
\node[hexa] (h7;4) at ({7/2+7/4},{(4+1/2)*sin(60)}) {51.9};

\node[hexa] (h5;2) at ({5/2+5/4},{(2+1/2)*sin(60)}) {2.6};
\node[hexa] (h7;2) at ({7/2+7/4},{(2+1/2)*sin(60)}) {2.6};
\node[hexa] (h5;5) at ({5/2+5/4},{(5+1/2)*sin(60)}) {2.6};
\node[hexa] (h7;5) at ({7/2+7/4},{(5+1/2)*sin(60)}) {2.6};

\node[hexa] (h3;4) at ({3/2+3/4},{(4+1/2)*sin(60)}) {5.5};
\node[hexa] (h9;4) at ({9/2+9/4},{(4+1/2)*sin(60)}) {5.5};
\node[hexa] (h3;3) at ({3/2+3/4},{(3+1/2)*sin(60)}) {5.5};
\node[hexa] (h9;3) at ({9/2+9/4},{(3+1/2)*sin(60)}) {5.5};

\node[hexa] (h3;5) at ({3/2+3/4},{(5+1/2)*sin(60)}) {0.3};
\node[hexa] (h9;5) at ({9/2+9/4},{(5+1/2)*sin(60)}) {0.3};
\node[hexa] (h3;2) at ({3/2+3/4},{(2+1/2)*sin(60)}) {0.3};
\node[hexa] (h9;2) at ({9/2+9/4},{(2+1/2)*sin(60)}) {0.3};

\node[hexa] (h1;4) at ({1/2+1/4},{(4+1/2)*sin(60)}) {0.1};
\node[hexa] (h11;4) at ({11/2+11/4},{(4+1/2)*sin(60)}) {0.1};
\node[hexa] (h1;3) at ({1/2+1/4},{(3+1/2)*sin(60)}) {0.1};
\node[hexa] (h11;3) at ({11/2+11/4},{(3+1/2)*sin(60)}) {0.1};

\node[hexa] (h4;2) at ({4/2+4/4},{2*sin(60)}) {0.1};
\node[hexa] (h4;6) at ({4/2+4/4},{6*sin(60)}) {0.1};
\node[hexa] (h8;2) at ({8/2+8/4},{2*sin(60)}) {0.1};
\node[hexa] (h8;6) at ({8/2+8/4},{6*sin(60)}) {0.1};

\node [circle,draw,red,thick,minimum size=1cm] at (h6;4){};

\end{tikzpicture}
\vspace{10pt}
\caption{Gaussian blur of a single bin (red circle) in the case $\sigma_x = 2$ and $\sigma_y = 1$ in units of hexagon side lengths. The numbers represent the transferred signal relative to the blurred value of the original bin (100\%). All other bins are below 0.1\% in this example.}
\label{fig4}
\end{center}
\end{figure}

We implemented a generic function combining Equations 2 and 3 for creating a blurred histogram data matrix in Python as part of SCOPE (Search and Chemical Ontology Plotting Environment). The SCOPE project includes tools for executing literature searches, collecting text-mined chemical entities of biological interest and creating interactive visualizations. The histograms were constructed using pandas and visualized by Bokeh version 1.4.0. Bokeh is particularly popular for creating interactive visualizations for web browsers and scales well to large datasets, including the millions of named entities that can be retrieved from a single literature search. The Bokeh interface in SCOPE includes sliders to adjust the blurring and scaling, and displays the most frequent chemical classes in any selected bin.

\pagebreak

\section{Verifications}
\label{sec:verify}

To visually verify that the Gaussian blurring does what it was designed to do, we can look at the effect of varying $\sigma_x$ and $\sigma_y$ in the kernel (Eq. 3):

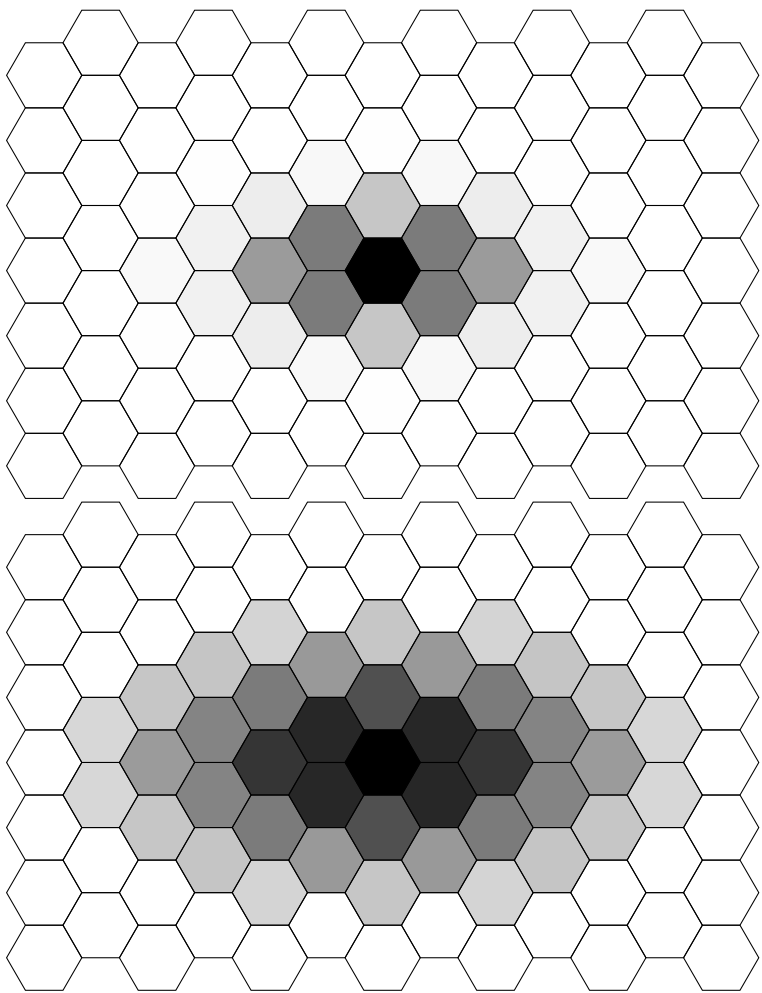
\begin{figure}[ht]
\begin{center}
\begin{tikzpicture} [hexa/.style= {shape=regular polygon,
                                   regular polygon sides=6,
                                   minimum size=1cm, 
                                   inner sep=0,anchor=south,
                                   fill=none}]
\fontsize{9}{11} 

\definecolor{gray1}{gray}{0}
\definecolor{gray0393}{gray}{0.607}
\definecolor{gray0024}{gray}{0.976}
\definecolor{gray0223}{gray}{0.777}
\definecolor{gray0002}{gray}{0.998}
\definecolor{gray0072}{gray}{0.928}
\definecolor{gray0519}{gray}{0.481}
\definecolor{gray0026}{gray}{0.974}
\definecolor{gray0055}{gray}{0.945}
\definecolor{gray0003}{gray}{0.997}
\definecolor{gray0001}{gray}{0.999}

\node[hexa,fill,gray0393,minimum size=1cm] (h4;4) at ({4/2+4/4},{4*sin(60)}) {39.3};
\node[hexa,fill,gray0393,minimum size=1cm] (h8;4) at ({8/2+8/4},{4*sin(60)}) {39.3};

\node[hexa,fill,gray0024,minimum size=1cm]  (h2;4) at ({2/2+2/4},{4*sin(60)}) {2.4};
\node[hexa,fill,gray0024,minimum size=1cm] (h10;4) at ({10/2+10/4},{4*sin(60)}) {2.4};

\node[hexa,fill,gray0223,minimum size=1cm] (h6;3) at ({6/2+6/4},{3*sin(60)}) {22.3};
\node[hexa,fill,gray0223,minimum size=1cm] (h6;5) at ({6/2+6/4},{5*sin(60)}) {22.3};

\node[hexa,fill,gray0002,minimum size=1cm] (h6;2) at ({6/2+6/4},{2*sin(60)}) {0.2};
\node[hexa,fill,gray0002,minimum size=1cm] (h6;6) at ({6/2+6/4},{6*sin(60)}) {0.2};

\node[hexa,fill,gray0072,minimum size=1cm] (h4;3) at ({4/2+4/4},{3*sin(60)}) {7.2};
\node[hexa,fill,gray0072,minimum size=1cm] (h4;5) at ({4/2+4/4},{5*sin(60)}) {7.2};
\node[hexa,fill,gray0072,minimum size=1cm] (h8;3) at ({8/2+8/4},{3*sin(60)}) {7.2};
\node[hexa,fill,gray0072,minimum size=1cm] (h8;5) at ({8/2+8/4},{5*sin(60)}) {7.2};

\node[hexa,fill,gray0519,minimum size=1cm] (h5;3) at ({5/2+5/4},{(3+1/2)*sin(60)}) {51.9};
\node[hexa,fill,gray0519,minimum size=1cm] (h5;4) at ({5/2+5/4},{(4+1/2)*sin(60)}) {51.9};
\node[hexa,fill,gray0519,minimum size=1cm] (h7;3) at ({7/2+7/4},{(3+1/2)*sin(60)}) {51.9};
\node[hexa,fill,gray0519,minimum size=1cm] (h7;4) at ({7/2+7/4},{(4+1/2)*sin(60)}) {51.9};

\node[hexa,fill,gray0026,minimum size=1cm] (h5;2) at ({5/2+5/4},{(2+1/2)*sin(60)}) {2.6};
\node[hexa,fill,gray0026,minimum size=1cm]  (h7;2) at ({7/2+7/4},{(2+1/2)*sin(60)}) {2.6};
\node[hexa,fill,gray0026,minimum size=1cm]  (h5;5) at ({5/2+5/4},{(5+1/2)*sin(60)}) {2.6};
\node[hexa,fill,gray0026,minimum size=1cm]  (h7;5) at ({7/2+7/4},{(5+1/2)*sin(60)}) {2.6};

\node[hexa,fill,gray0055,minimum size=1cm] (h3;4) at ({3/2+3/4},{(4+1/2)*sin(60)}) {5.5};
\node[hexa,fill,gray0055,minimum size=1cm] (h9;4) at ({9/2+9/4},{(4+1/2)*sin(60)}) {5.5};
\node[hexa,fill,gray0055,minimum size=1cm] (h3;3) at ({3/2+3/4},{(3+1/2)*sin(60)}) {5.5};
\node[hexa,fill,gray0055,minimum size=1cm] (h9;3) at ({9/2+9/4},{(3+1/2)*sin(60)}) {5.5};

\node[hexa,fill,gray0003,minimum size=1cm] (h3;5) at ({3/2+3/4},{(5+1/2)*sin(60)}) {0.3};
\node[hexa,fill,gray0003,minimum size=1cm] (h9;5) at ({9/2+9/4},{(5+1/2)*sin(60)}) {0.3};
\node[hexa,fill,gray0003,minimum size=1cm] (h3;2) at ({3/2+3/4},{(2+1/2)*sin(60)}) {0.3};
\node[hexa,fill,gray0003,minimum size=1cm] (h9;2) at ({9/2+9/4},{(2+1/2)*sin(60)}) {0.3};

\node[hexa,fill,gray0001,minimum size=1cm] (h1;4) at ({1/2+1/4},{(4+1/2)*sin(60)}) {0.1};
\node[hexa,fill,gray0001,minimum size=1cm] (h11;4) at ({11/2+11/4},{(4+1/2)*sin(60)}) {0.1};
\node[hexa,fill,gray0001,minimum size=1cm] (h1;3) at ({1/2+1/4},{(3+1/2)*sin(60)}) {0.1};
\node[hexa,fill,gray0001,minimum size=1cm] (h11;3) at ({11/2+11/4},{(3+1/2)*sin(60)}) {0.1};

\node[hexa,fill,gray0001,minimum size=1cm] (h4;2) at ({4/2+4/4},{2*sin(60)}) {0.1};
\node[hexa,fill,gray0001,minimum size=1cm] (h4;6) at ({4/2+4/4},{6*sin(60)}) {0.1};
\node[hexa,fill,gray0001,minimum size=1cm] (h8;2) at ({8/2+8/4},{2*sin(60)}) {0.1};
\node[hexa,fill,gray0001,minimum size=1cm] (h8;6) at ({8/2+8/4},{6*sin(60)}) {0.1};

\node[hexa,fill,gray0002,minimum size=1cm] (h2;3) at ({2/2+2/4},{3*sin(60)}) {0.2}; %
\node[hexa,fill,gray0002,minimum size=1cm] (h10;3) at ({10/2+10/4},{3*sin(60)}) {0.2}; %
\node[hexa,fill,gray0002,minimum size=1cm] (h2;5) at ({2/2+2/4},{5*sin(60)}) {0.2}; %
\node[hexa,fill,gray0002,minimum size=1cm] (h10;5) at ({10/2+10/4},{5*sin(60)}) {0.2}; %

\node[hexa,fill,gray1,minimum size=1cm] (h6;4) at ({6/2+6/4},{4*sin(60)}) {100};

\foreach \j in {0,...,12}{ 
    \ifodd\j 
       \foreach \i in {1,...,7}{\node[hexa,draw] (h\j;\i) at ({\j/2+\j/4},{(\i+1/2)*sin(60)}){};}        
    \else
        \foreach \i in {1,...,7}{\node[hexa,draw] (h\j;\i) at ({\j/2+\j/4},{\i*sin(60)}){};}
    \fi}

\end{tikzpicture}
\begin{tikzpicture} [hexa/.style= {shape=regular polygon,
                                   regular polygon sides=6,
                                   minimum size=1cm, 
                                   inner sep=0,anchor=south,
                                   fill=none}]
\fontsize{9}{11} 
\definecolor{gray1}{gray}{0}
\definecolor{gray0792}{gray}{0.208} %
\definecolor{gray0393}{gray}{0.607} %
\definecolor{gray0687}{gray}{0.313} %
\definecolor{gray0223}{gray}{0.777} %
\definecolor{gray0519}{gray}{0.481} %
\definecolor{gray0849}{gray}{0.151} %
\definecolor{gray0401}{gray}{0.599} %
\definecolor{gray0484}{gray}{0.516} %
\definecolor{gray0228}{gray}{0.772} %
\definecolor{gray0157}{gray}{0.843} %
\definecolor{gray0168}{gray}{0.832} %

\node[hexa,fill,gray0792,minimum size=1cm] (h4;4) at ({4/2+4/4},{4*sin(60)}) {79.2}; %
\node[hexa,fill,gray0792,minimum size=1cm] (h8;4) at ({8/2+8/4},{4*sin(60)}) {79.2}; %

\node[hexa,fill,gray0393,minimum size=1cm]  (h2;4) at ({2/2+2/4},{4*sin(60)}) {39.3}; %
\node[hexa,fill,gray0393,minimum size=1cm] (h10;4) at ({10/2+10/4},{4*sin(60)}) {39.3}; %

\node[hexa,fill,gray0687,minimum size=1cm] (h6;3) at ({6/2+6/4},{3*sin(60)}) {68.7}; %
\node[hexa,fill,gray0687,minimum size=1cm] (h6;5) at ({6/2+6/4},{5*sin(60)}) {68.7}; %

\node[hexa,fill,gray0223,minimum size=1cm] (h6;2) at ({6/2+6/4},{2*sin(60)}) {22.3}; %
\node[hexa,fill,gray0223,minimum size=1cm] (h6;6) at ({6/2+6/4},{6*sin(60)}) {22.3}; %

\node[hexa,fill,gray0519,minimum size=1cm] (h4;3) at ({4/2+4/4},{3*sin(60)}) {51.9}; %
\node[hexa,fill,gray0519,minimum size=1cm] (h4;5) at ({4/2+4/4},{5*sin(60)}) {51.9}; %
\node[hexa,fill,gray0519,minimum size=1cm] (h8;3) at ({8/2+8/4},{3*sin(60)}) {51.9}; %
\node[hexa,fill,gray0519,minimum size=1cm] (h8;5) at ({8/2+8/4},{5*sin(60)}) {51.9}; %

\node[hexa,fill,gray0849,minimum size=1cm] (h5;3) at ({5/2+5/4},{(3+1/2)*sin(60)}) {84.9}; %
\node[hexa,fill,gray0849,minimum size=1cm] (h5;4) at ({5/2+5/4},{(4+1/2)*sin(60)}) {84.9}; %
\node[hexa,fill,gray0849,minimum size=1cm] (h7;3) at ({7/2+7/4},{(3+1/2)*sin(60)}) {84.9}; %
\node[hexa,fill,gray0849,minimum size=1cm] (h7;4) at ({7/2+7/4},{(4+1/2)*sin(60)}) {84.9}; %

\node[hexa,fill,gray0401,minimum size=1cm] (h5;2) at ({5/2+5/4},{(2+1/2)*sin(60)}) {40.1}; %
\node[hexa,fill,gray0401,minimum size=1cm]  (h7;2) at ({7/2+7/4},{(2+1/2)*sin(60)}) {40.1}; %
\node[hexa,fill,gray0401,minimum size=1cm]  (h5;5) at ({5/2+5/4},{(5+1/2)*sin(60)}) {40.1}; %
\node[hexa,fill,gray0401,minimum size=1cm]  (h7;5) at ({7/2+7/4},{(5+1/2)*sin(60)}) {40.1}; %

\node[hexa,fill,gray0484,minimum size=1cm] (h3;4) at ({3/2+3/4},{(4+1/2)*sin(60)}) {48.4}; %
\node[hexa,fill,gray0484,minimum size=1cm] (h9;4) at ({9/2+9/4},{(4+1/2)*sin(60)}) {48.4}; %
\node[hexa,fill,gray0484,minimum size=1cm] (h3;3) at ({3/2+3/4},{(3+1/2)*sin(60)}) {48.4}; %
\node[hexa,fill,gray0484,minimum size=1cm] (h9;3) at ({9/2+9/4},{(3+1/2)*sin(60)}) {48.4}; %

\node[hexa,fill,gray0228,minimum size=1cm] (h3;5) at ({3/2+3/4},{(5+1/2)*sin(60)}) {22.8}; %
\node[hexa,fill,gray0228,minimum size=1cm] (h9;5) at ({9/2+9/4},{(5+1/2)*sin(60)}) {22.8}; %
\node[hexa,fill,gray0228,minimum size=1cm] (h3;2) at ({3/2+3/4},{(2+1/2)*sin(60)}) {22.8}; %
\node[hexa,fill,gray0228,minimum size=1cm] (h9;2) at ({9/2+9/4},{(2+1/2)*sin(60)}) {22.8}; %

\node[hexa,fill,gray0157,minimum size=1cm] (h1;4) at ({1/2+1/4},{(4+1/2)*sin(60)}) {15.7}; %
\node[hexa,fill,gray0157,minimum size=1cm] (h11;4) at ({11/2+11/4},{(4+1/2)*sin(60)}) {15.7}; %
\node[hexa,fill,gray0157,minimum size=1cm] (h1;3) at ({1/2+1/4},{(3+1/2)*sin(60)}) {15.7}; %
\node[hexa,fill,gray0157,minimum size=1cm] (h11;3) at ({11/2+11/4},{(3+1/2)*sin(60)}) {15.7}; %

\node[hexa,fill,gray0168,minimum size=1cm] (h4;2) at ({4/2+4/4},{2*sin(60)}) {16.8}; %
\node[hexa,fill,gray0168,minimum size=1cm] (h4;6) at ({4/2+4/4},{6*sin(60)}) {16.8}; %
\node[hexa,fill,gray0168,minimum size=1cm] (h8;2) at ({8/2+8/4},{2*sin(60)}) {16.8}; %
\node[hexa,fill,gray0168,minimum size=1cm] (h8;6) at ({8/2+8/4},{6*sin(60)}) {16.8}; %

\node[hexa,fill,gray0223,minimum size=1cm] (h2;3) at ({2/2+2/4},{3*sin(60)}) {22.3}; %
\node[hexa,fill,gray0223,minimum size=1cm] (h10;3) at ({10/2+10/4},{3*sin(60)}) {22.3}; %
\node[hexa,fill,gray0223,minimum size=1cm] (h2;5) at ({2/2+2/4},{5*sin(60)}) {22.3}; %
\node[hexa,fill,gray0223,minimum size=1cm] (h10;5) at ({10/2+10/4},{5*sin(60)}) {22.3}; %

\node[hexa,fill,gray1,minimum size=1cm] (h6;4) at ({6/2+6/4},{4*sin(60)}) {100};

\foreach \j in {0,...,12}{ 
    \ifodd\j 
       \foreach \i in {1,...,7}{\node[hexa,draw] (h\j;\i) at ({\j/2+\j/4},{(\i+1/2)*sin(60)}){};}        
    \else
        \foreach \i in {1,...,7}{\node[hexa,draw] (h\j;\i) at ({\j/2+\j/4},{\i*sin(60)}){};}
    \fi}

\end{tikzpicture}
\caption{Examples of a single blurred hexagonal bin using the numbers in Figure 4 (top) and with $\sigma_x = 4$ and $\sigma_y = 2$ (bottom), restricting the blur to the same cells. The color mapping in both cases is linear from white at 0\% to black at 100\% of the blurred value of the central bin.}
\label{fig5} 
\end{center}
\end{figure}

We tested our Gaussian blurring implementation for Bokeh using SCOPE and searching Europe PMC for scientific papers mentioning particular techniques from analytical chemistry in their section-tagged materials and methods sections \citep{palmblad19}. Hexagonal bins more faithfully traced the shapes of log \textit{P}/mass distributions of chemical classes and the resulting histograms exhibited different distributions as expected from what is known about the applicability of the analytical techniques:

\begin{figure}[ht]
\begin{center}
\includegraphics[width=0.49\textwidth]{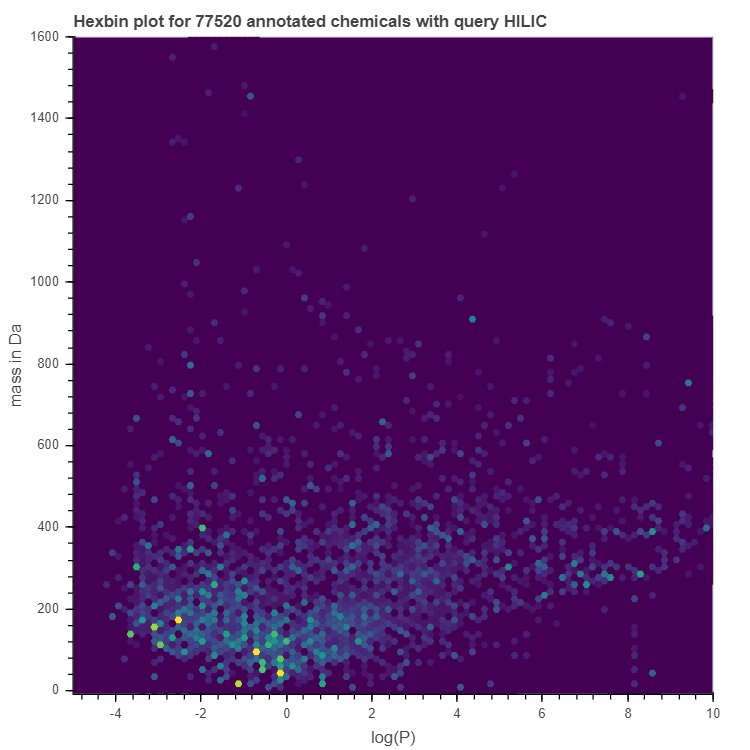}
\includegraphics[width=0.49\textwidth]{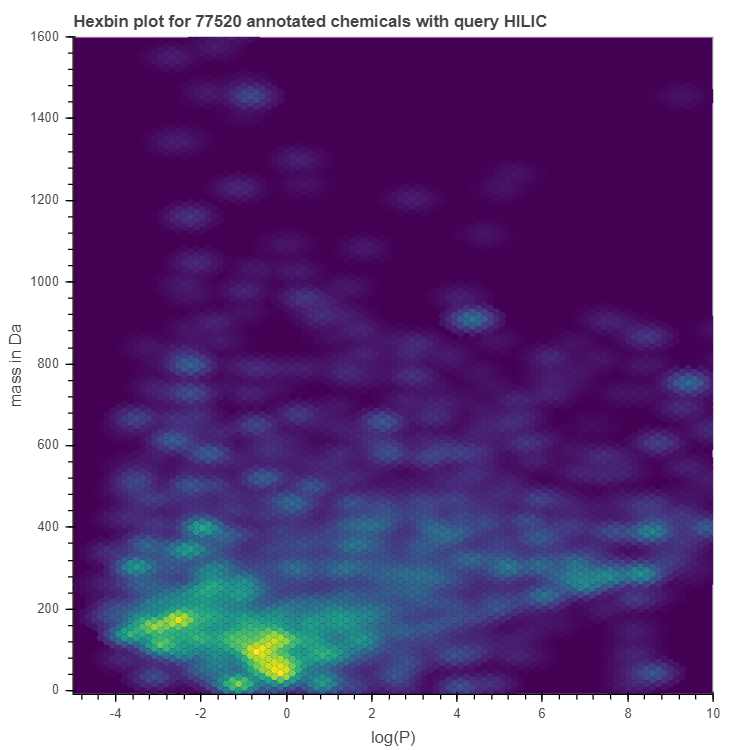}
\caption{Distributions of mass (Da) and polarity (log \textit{P}) of small molecules retrieved from papers in Europe PMC mentioning hydrophilic interaction chromatography (HILIC) in their methods section. The distributions are visualized by SCOPE before (left) and after (right) applying a non-uniform Gaussian blur of size 2.5. The saturation was set to 2.5, term frequency-inverse document frequency normalization applied and the Viridis colormap selected.}
\label{fig6}
\end{center}
\end{figure}

The sliders for adjusting the blurring and saturation improve the user experience and help create clear visualizations. The blurring convey the uncertainty in the log \textit{P} predictions. We also found interactive bin selection to be easier in hexagonal bins than in equiareal rectangular bins. Other features and applications of SCOPE are beyond the scope of the this paper (no pun intended), but will be described elsewhere.

\section{Conclusion}
\label{sec:conc}

In this short paper we have shown how to calculate and apply non-uniform Gaussian blurs in Cartesian coordinates in hexagonal bins, and illustrated this with a real-world example. For simplicity, we evaluated the Gaussian kernel in the center of each hexagonal bin. It may be argued that the \textit{average} kernel value in the bin more accurate represents the underlying distribution. However, this would require numerical integration over all these hexagonal bins for what would likely be a very small visual or practical benefit. All source code is available on GitHub (ReinV/SCOPE) under the Apache 2.0 license.

\bibliographystyle{chicago}
\bibliography{library}
\end{document}